\documentstyle[preprint,aps,floats,psfig]{revtex}
\begin{document}

\draft

\title{On the decay of localized vibrational states in glasses: a
one-dimensional example} 
\author{Jaroslav Fabian}
\address{Department of Physics, State University of New York at Stony Brook, 
Stony Brook, New York 11794-3800}
\maketitle

\vspace{1cm}

\begin{abstract}

The interaction between three localized vibrational modes 
is shown to be as relevant for the lifetimes of localized
modes as the interaction involving two localized and one 
extended, and one localized and two extended modes. This 
contrasts with previous views. I support my arguments by
a numerical study of a strongly disordered linear atomic 
chain.

\end{abstract}

\vspace{2em}

\newpage

Lack of a periodic structure makes the character of the 
vibrational states (VS) in glasses rather complex.
Their spatio-temporal properties depend on the spectral
region of interest. The lowest frequency VS are 
propagons ($p$, terminology from \cite{fabian}) - sound 
waves with a wavevector (of magnitude $\rm 2 \pi/\lambda$), 
polarization and velocity.
They elastically scatter off local structural imperfections,
and as long as their mean free path $\rm  \ell >\lambda$,
the scattering does not destroy their wave character.
What happens at the Ioffe-Regel frequency 
\cite{ioffe} $\omega_{\rm IR}$,
where $\rm \ell \approx \lambda$ (and the wavevector concept becomes
invalid) is still a matter of debate.
Since $\omega_{\rm IR}$ is very low (less than 5 meV for amorphous
silicon (a-Si) where the maximum frequency is $\approx$ 80 meV), 
VS with $\omega\agt\omega_{\rm IR}$ are responsible for thermodynamic
and transport properties of glasses at temperatures ranging
from several Kelvins up to the melting point.

Presently there are
two different scenarios for what happens beyond the Ioffe-Regel limit.
(i) The ``fracton'' model \cite{orbach} postulates that all VS
with $\omega\agt\omega_{\rm IR}$ are ``locons'' ($l$), localized
modes which are anderson-localized \cite{anderson} because of 
topological disorder. The relation to ``fractal'' geometry is
an historical accident.
(ii) In the ``diffuson'' model 
\cite{allen1,allen2} the majority of modes are ``diffusons'', 
extended and non-propagating modes, carrying energy diffusively.
Numerical simulations on realistic models \cite{allen2,local} show 
that locons form only a small portion ($\rm \sim 3\%$ in a-Si) of 
the highest-frequency modes.

In harmonic approximation the VS will never equilibrate, 
but anharmonic interactions are always present. What is the
rate of equilibration in glasses? Experiment  
\cite{scholten} indicates that at
a liquid He temperature in a-Si 
(a) high frequency ($\omega \agt$ 10 meV) VS live longer 
as frequency  increases, and 
(b) the VS live on nanosecond timescales. In contrast, phonons in crystals 
decay faster as frequency increases
(the number of decay channels increases), and lifetimes are  picoseconds.
Model (i) seems to agree with
the experiment in the following way. Locons $l$ can decay via three processes:
(1) $l\rightarrow p'p''$, 
(2) $l\rightarrow p'l''$ or $lp'\rightarrow l''$,
and (3) $l\rightarrow l'l''$ or $ll'\rightarrow l''$.
Process (1) is kinematically forbidden for high
frequency locons with $\omega_l
> 2\omega_{\rm IR}$. Processes (2) represent ``propagon-assisted hopping''
between locons $l$ and $l''$. Because $\omega_{p'} \ll (\omega_l,
\omega_{l''})$, 
$\omega_l\approx\omega_{l''}$, and the locons $l$ and $l''$ 
spatially repel each other \cite{halperin}. As the frequency 
and the inverse localization legth $\rm L^{-1}$ increases, 
the overlap between the locons decreases
and so does the hopping rate.
Processes (3) have been neglected by model (i) on the grounds 
of a small  probability of three locon overlap. Hopping is 
therefore assumed to be the only locon decay process, and because it has
property (a) and can be ``adjusted'' to fit (b), the experiment
\cite{scholten} seems to be rationalized. 

Recent numerical calculations on models (i) and (ii) 
\cite{fabian} have shown that decay properties of both diffusons 
and locons are similar to those of  crystalline phonons. 
It has been argued \cite{fabian} that the assumption of the negligibility
of (3) is incorrect and is the reason behind the discrepancy between
the numerical results and model (i) predictions. The experiment
\cite{scholten} therefore remains to be explained.
In this paper I  show, using  a one-dimensional (1d)
numerical realization of model (i), that extended and localized
modes decay on the same time scale and that the three locon
interaction is dominant in the decay of locons with $\omega \agt 2\omega_c$.

Consider a linear chain of N uniformly spaced atoms connected by 
random springs \cite{dyson,kori}.  
Periodic boundary conditions are assumed.
In terms of displacements $u_a$ ($ u_{\rm N+1}\equiv u_{\rm 1}$) of 
atoms $a$ from equilibrium,
the potential energy of the system can be expressed as
\begin{eqnarray} \label{eqn:potential}
V=\frac{1}{2} \sum_{a=1}^{N} K_{a} (u_a-u_{a+1})^2.
\end{eqnarray}
Here $ K_{a}$=$ K_0(1+\xi_{a})$, where 
$\xi_{a}$ are random numbers uniformly distributed 
in the interval $<$-b,b$>$, and
$ K_0$=10.6 $\rm eV/\AA^2$ is chosen to simulate a 
hypothetical linear silicon chain (the silicon atom
mass M) with maximum vibrational
frequency $\omega_{\rm max}=2\sqrt{K_0/M} \approx$ 80 meV in 
the case of no disorder (b=0). 
The length scale is the interatomic
spacing, the exact value of which is not important here. 
For the case N=3000, vibrational eigenstates and frequencies
were found by exact numerical diagonalization. 
This number is sufficiently large to ensure that the conclusions  
will not be affected by finite size effects. 
The goal is to numerically realize the ``fracton model" 
scenario with majority of VS localized. As shown later, 
the strong disorder value b=0.7 suits this purpose and 
will be used from now on. The ``clean'' case, b=0, will
serve as a reference.

Figure \ref{fig:1} shows vibrational density of states (DOS)
for the above model and for the case b=0. 
The latter can be trivially solved by introducing wavevectors.
Consequently its DOS has the $1/\sqrt{\omega^2_{\rm max}-\omega^2}$
dependence and a Van Hove singularity at $\omega_{\rm max}$.
As the disorder increases, the description of VS
in terms of wavevectors is less and less valid. 
For b=0.7, the Van Hove singularity  is washed out and DOS becomes
nearly flat.

In a 1d infinite system any disorder causes all VS 
to be localized \cite{mott}, with  
a trivial exemption for
the zero frequency mode corresponding to a translation 
of the system as a whole. However, if N is finite there are
always low frequency modes with L larger than the system size. 
They appear to be extended, sound wave-like modes, although
on intermediate timescales they behave as diffusons \cite{jon}.
Only after the system size is increased, they become manifestly
localized. 
A good measure of the number of atoms participating 
in the vibration of mode $i$ is the participation ratio $P_i$.
Its inverse is defined as
\begin{eqnarray}\label{eqn:participation}
1/P_i=\sum_{a=1}^N (e^i_a)^4, 
\end{eqnarray}
where $e^i_a$ are normalized ($ \sum_a (e^i_a)^2$ = 1)
vibrational eigenstates with frequencies $\omega_i$.
I plot 1/P in Fig. \ref{fig:2}.
1/P grows monotonously with frequency, the dependence being 
close to quadratic for $\omega \agt$ 10 meV. At lower frequencies
1/P $\approx$ 1/N, since L $\agt$ system size (for infinite N it
has been predicted \cite{azbel} that 1/P $\sim \omega^2$ as a 
result of 1d elastic sound wave scattering).
To accurately locate
the mobility edge $\omega_c$ I use the Thouless criterion \cite{thouless},
that  $\rm \Delta \omega/\delta \omega$ should exhibit a sharp drop
at $\omega_c$.
For a given mode, $\rm \Delta \omega$ is the locally averaged change 
of the mode 
frequency under the change of boundary conditions from periodic 
to antiperiodic, and $\rm \delta \omega$ is the local average level spacing.
I find it convenient to consider the cumulative quantity
$\tau(\omega)=\int_0^{\omega}d\omega'\Delta \omega(\omega')
/\delta \omega(\omega')$, which is plotted
in the inset of Fig. \ref{fig:2}. The mobility edge (vertical line) 
is $\omega_c \approx$ 9 meV and $\rm \Delta \omega/\delta \omega$ 
at $\omega_c$ is less than 1\%. Locons with $\omega \agt \omega_c$
form $\approx 90\%$ of the spectrum, enough to 
simulate model (i).

To get a qualitative understanding of how VS decay, consider
a simplified version of zero-temperature decay rate formula 
(full formula see e.g. \cite{maradudin}):
\begin{eqnarray} \label{eqn:overlap}
\tilde{\Gamma}_i=\sum_{j,k} (J_{ijk})^2
\delta(\omega_i-\omega_j-\omega_k),
\end{eqnarray}
where the sum is over all modes $j$ and $k$, and
\begin{eqnarray} \label{eqn:overlap1}
J_{ijk}= \sum_a \eta_a e^i_ae^j_ae^k_a
\end{eqnarray}
measures the overlap between modes $i$, $j$, and $k$.
$\eta_a$ is a binary random variable having values 1 or -1.
Only ``fission'' decay processes ($i\rightarrow jk$) are allowed
at zero-temperature. ``Fusion''  processes ($ij\rightarrow k$)
appearing at finite temperatures will not be considered here,
since they can be handled analogously.
In 1d VS have a partial memory of a wavevector, and therefore
a partial phase coherence. The reason is that the n-th vibrational eigenstate
has n-1 nodes (if the zero-frequency mode is n=1) \cite{dyson}.
$\eta_a$ is inserted into Eq. \ref{eqn:overlap1} to eliminate 
this coherence, so that the results will not be substantially affected
by dimensionality (apart from the DOS effects). Imagine 
VS with amplitudes independent of frequency. Then an expression 
similar to Eq. \ref{eqn:overlap}
can be obtained for the decay of the VS caused by a set of
nonlinear springs with potentials $\sim \eta_a u^3_a$ + O($u^4_a$), 
externally attached to each atom.

The spectral dependence of $\tilde{\Gamma}$ is shown
in Fig. \ref{fig:3}. The case b=0, plotted in (b), does
not represent the decay of 1d crystalline phonons, since the
phases of these are randomized by $\eta_a$ in Eq. \ref{eqn:overlap1}.
It, however, shows the hypothetical behavior of random-phase $extended$
modes, such as occur below the mobility edge in 3d. The $\omega$-variation
of $\tilde{\Gamma}$ follows the joint density of states (JDOS),  
$\sum_{jk}\delta(\omega_i-\omega_j-\omega_k)$,
and provides a benchmark when discussing the decay behavior of
locons.  Data in (a) at $\omega\agt\omega_c$ are much more 
scattered than in (b), since the locon decay involves some 
overlap statistics (explained below).
Therefore local averages are used to represent the data. 
$\tilde{\Gamma}$ in (a) increases with frequency and follows
JDOS, having similar magnitudes as in (b). According to
model (i) there would be a sharp decrease in $\tilde{\Gamma}$
beyond 2$\omega_c$.
This certainly does not happen.

The reason why the behavior of $\tilde{\Gamma}$ differs from the
one predicted by model (i) is that processes (3) can not
be neglected. In fact they are dominant decay processes for high
frequency locons. Consider a decay $i\rightarrow jk$, where $i$, $j$, 
and $k$
are locons with corresponding participation numbers $ P_i$, $ P_j$, 
and $P_k$. Eq. \ref{eqn:overlap} can be rewritten as
$ \tilde{\Gamma}_i =\int\int d\omega_j d\omega_k N(\omega_j)
N(\omega_k) J^2_i(\omega_j,\omega_k) \delta(\omega_i-\omega_j-\omega_k)$,
where $N(\omega_j)$ is the total number of states at $\omega_j$, and
$ J^2_i(\omega_j,\omega_k)$ is $ J^2_{ijk}$ averaged over frequency
shells  of $\omega_j$ and $\omega_k$.
A simple scaling argument \cite{fabian} shows that
$ J^2_i(\omega_j,\omega_k)$ does not explicitly depend on $ P_i$,
$ P_j$, or $ P_k$. It goes as follows. The maximum decay magnitude
among $ J^2_{ijk}$ is found  when $i$ lies entirely within an overlap of $j$ 
and $k$, and  scales as $ 1/P_jP_k$ 
($ e^i_a\sim 1/\sqrt{P_i}$, etc., and the sum in 
Eq. \ref{eqn:overlap1} is over $P_i$ random-sign numbers
and scales as $\sqrt{P_i}$). The probability for this overlap
to occur scales as $ P_jP_k/{\rm N^2}$.  
An important assumption used here is that locons $i$, $j$, and $k$, having
different frequencies, are spatially uncorrelated (illustrated below).
The average $ J^2_i(\omega_j,\omega_k)$ therefore scales as $\rm 1/N^2$ and 
$ \tilde{\Gamma}_i$ is N-independent (and follows the JDOS). 
The small probability
of the three-locon overlap is exactly compensated by the large magnitudes
of decay matrix elements, when the overlap occurs. 
As for  hopping processes $l\rightarrow p'l''$,
one has to distinguish between two cases. If $\omega_{p'} \ll 
\omega_l,\omega_{l''}$ (as in model (i) and this paper), the two 
locons $l$ and $l''$ spatially repel  each other  \cite{halperin}
and the decay rates decrease as L of $l$ and $l''$ decreases.  
On the other
hand, if $\omega_{p'}$ is not small, one can apply the 
above scaling arguments,
since majority of $l$ and $l''$ are not spatially
correlated. Decay rates of this 
``uncorrelated'' hopping are therefore independent of the localization
character of VS involved and follow the JDOS. 
When a low frequency ($ \omega_l \alt {\rm 2}\omega_c$) locon decays into two
propagons, $l\rightarrow p'p''$, the overlap probability is one
($P_j, P_k$ = N) and the case is trivial, similar in behavior
to three locon and ``uncorrelated'' hopping processes.
These arguments do not
depend on d and worked, as far as one could tell, in 3d also \cite{fabian}.

Figure \ref{fig:4} illustrates the foregoing discussion. Data are 
represented by running averages. Decay $l\rightarrow l'l''$ in (a),
obtained by considering only j and k with $\omega_j$, $\omega_k \ge \omega_c$ in
Eq. \ref{eqn:overlap},
dominates the high frequency region and grows monotonously with frequency.
When, on the other hand, restricting, say modes j to be propagons 
($\omega_j \le \omega_c$), one gets hopping $l\rightarrow p'l''$ which 
decreases as frequency is increased. 
Since $\omega_c$ is large enough to allow 
hopping between locons from different spectral regions, the
decrease is quite moderate. The graph (b) in Fig. \ref{fig:4} shows
hopping when propagon frequency is restricted  by
$ \omega_p \le$ 1 meV. Computed rates are scattered over several 
magnitudes. In a sample of decay rates of locons confined into 
a small frequency band (of the width, say, 1 meV), majority 
have negligible values, but there are $\alt$ 10\% of large
magnitude rates giving rise to large local averages (they come
from a residual overlap between locons with similar frequencies). 
As the graph shows, local medians sharply decrease with frequency
and deviate strongly from local averages.
Since the local distribution of decay rates is found to have a strong central
tendency around small values (the area in the tails is small), the
medians better represent computed data. For other decay
types local medians and averages are similar.

Finally, I present figure \ref{fig:5} to  support the  argument 
that locons
from different spectral regions are spatially uncorrelated, as
opposed to locons with similar frequencies, which repel each other.
N=1000 is used for this purpose. The spectrum is divided
into $\delta$=1 meV wide bands, each band containing 10 VS on
the average. For every mode $i$ in a band only atoms $a$ having
$(e^i_a)^2\ge$ 0.2 $(e^i_{\rm max})^2$ are shown, $e^i_{\rm max}$
being the largest magnitude found among the coordinates of the vector
${\bf e}^i$. Localization starts somewhere beyond 15 meV,
where fewer than P $\approx$ 200 atoms participate in the vibration. 
There is clearly no statistically relevant overlap between the locons
lying within a frequency band. On the other hand, locons from different
spectral regions show no sign of repulsion, their spatial
positions are essentially  uncorrelated.

In summary, I have used a 1d random spring atomic model to
demonstrate that the interaction including three locons is
important for the locon decay processes. I have shown that 
extended mode and locon decay rates have similar magnitudes,
increasing with increasing frequency and closely following JDOS.
Hopping decay rates decrease as frequency increases, the decrease
is sharper as the frequency difference of hopping locons gets smaller.

I thank P. B. Allen for useful comments on the manuscript, and 
J. L. Feldman and S. Bickham for stimulating 
discussions. This work was supported by NSF Grant No. DMR 9417755.

\newpage

\begin{figure}
\caption{Calculated DOS for the 3000 atom
linear chain with the spring constant disorder of b=0.7 (solid
line) and b=0 (dashed line).
}
\label{fig:1}
\end{figure}

\begin{figure}
\caption{Inverse participation ratio 1/P and the cumulative Thouless
number $\tau$ (inset) as a function of frequency. The vertical line
in the inset is the mobility edge $\omega_c\approx$ 9 meV.
}
\label{fig:2}
\end{figure}

\begin{figure}
\caption{Calculated $\tilde{\Gamma}$ (circles represent
uniformly sampled 1/10-th of data) for b=0.7 (a) and b=0 (b).
Scattered data in (a) are represented by their running averages
(solid line). Vertical lines indicate $\omega_c$ and 2$\omega_c$,
and dashed lines are JDOS. Units are arbitrary, though the same in
(a) and (b).}
\label{fig:3}
\end{figure}

\begin{figure}
\caption{Calculated $\tilde{\Gamma}$ (a) for
the three locon (solid) and hopping (dashed) decay processes,
and (b) for hopping between locons from the same frequency region
(circles - see caption to Fig. \protect\ref{fig:3}).
Vertical line in (a) indicates $\omega_c$.
Units are the same as in Fig. \protect\ref{fig:3}.
}
\label{fig:4}
\end{figure}

\begin{figure}
\caption{
Localization of VS in the 1000 atom linear random spring model.
For each mode only the atoms vibrating with amplitudes higher than
a certain value (see text) are shown. The solid line is the particiation
ratio P.
}
\label{fig:5}
\end{figure}

\end{document}